\documentclass[reprint, amsmath,amssymb, aps,prb,
superscriptaddress,nofootinbib]{revtex4-2}
\usepackage{graphicx}
\usepackage{color}

\newcommand{\MSD}{{\rm MSD}}
\begin{document}
\title{Anomalous quantum transport in fractal lattices}

\author{Abel Rojo-Francàs}
\affiliation{DIPC - Donostia International Physics Center, Paseo Manuel de Lardiz{\'a}bal 4, 20018 San Sebasti{\'a}n, Spain}
\affiliation{Departament de F{\'i}sica Qu{\`a}ntica i Astrof{\'i}sica, Facultat de F{\'i}sica, Universitat de Barcelona, E-08028 Barcelona, Spain}
\affiliation{Institut de Ci{\`e}ncies del Cosmos, Universitat de Barcelona, ICCUB, Mart{\'i} i Franqu{\`e}s 1, E-08028 Barcelona, Spain.}

\author{Priyanshu Pansari}
\affiliation{Indian Institute of Technology, Roorkee, India}

\author{Utso Bhattacharya}
\affiliation{ICFO - Institut de Ci\`encies Fot\`oniques, The Barcelona Institute of Science and Technology, 08860 Castelldefels (Barcelona), Spain}
\affiliation{Institute for Theoretical Physics, ETH Zurich, 8093 Zurich, Switzerland}

\author{Bruno Juliá-Díaz}
\affiliation{Departament de F{\'i}sica Qu{\`a}ntica i Astrof{\'i}sica, Facultat de F{\'i}sica, Universitat de Barcelona, E-08028 Barcelona, Spain}
\affiliation{Institut de Ci{\`e}ncies del Cosmos, Universitat de Barcelona, ICCUB, Mart{\'i} i Franqu{\`e}s 1, E-08028 Barcelona, Spain.}

\author{Tobias Grass}
\affiliation{DIPC - Donostia International Physics Center, Paseo Manuel de Lardiz{\'a}bal 4, 20018 San Sebasti{\'a}n, Spain}
\affiliation{IKERBASQUE, Basque Foundation for Science, Plaza Euskadi 5, 48009 Bilbao, Spain}


\begin{abstract}
Fractal lattices are self-similar structures with repeated patterns on different scales. As in other aperiodic lattices, the absence of translational symmetry can give rise to quantum localization effects. In contrast to low-dimensional disordered systems, co-existence of localized and extended states is possible in fractal structures, and can lead to subtle transport behavior. Here, we study the dynamical properties of two fractal lattices, the Sierpiński gasket and the Sierpiński carpet. Despite their geometric similarity, the transport turns out to behave very differently: In the Sierpiński gasket, we find a sub-diffusive behavior, whereas the Sierpiński carpet exhibits sub-ballistic transport properties. We show that the different dynamical behavior is in line with qualitative differences of the systems' spectral properties. Specifically, in contrast to the Sierpiński carpet, the Sierpiński gasket exhibits an inverse power-law behavior of the level spacing distribution. From the point of view of technological applications, we demonstrate that the sub-diffusive behavior in the Sierpiński gasket can be used as a quantum memory. By interpolating between fractal and regular lattices, a flexible tuning between different transport regimes becomes possible.
\end{abstract}

\maketitle   

\section{Introduction}\label{sec:intro}
\begin{figure}
\centering
    \includegraphics[width=0.48\textwidth]{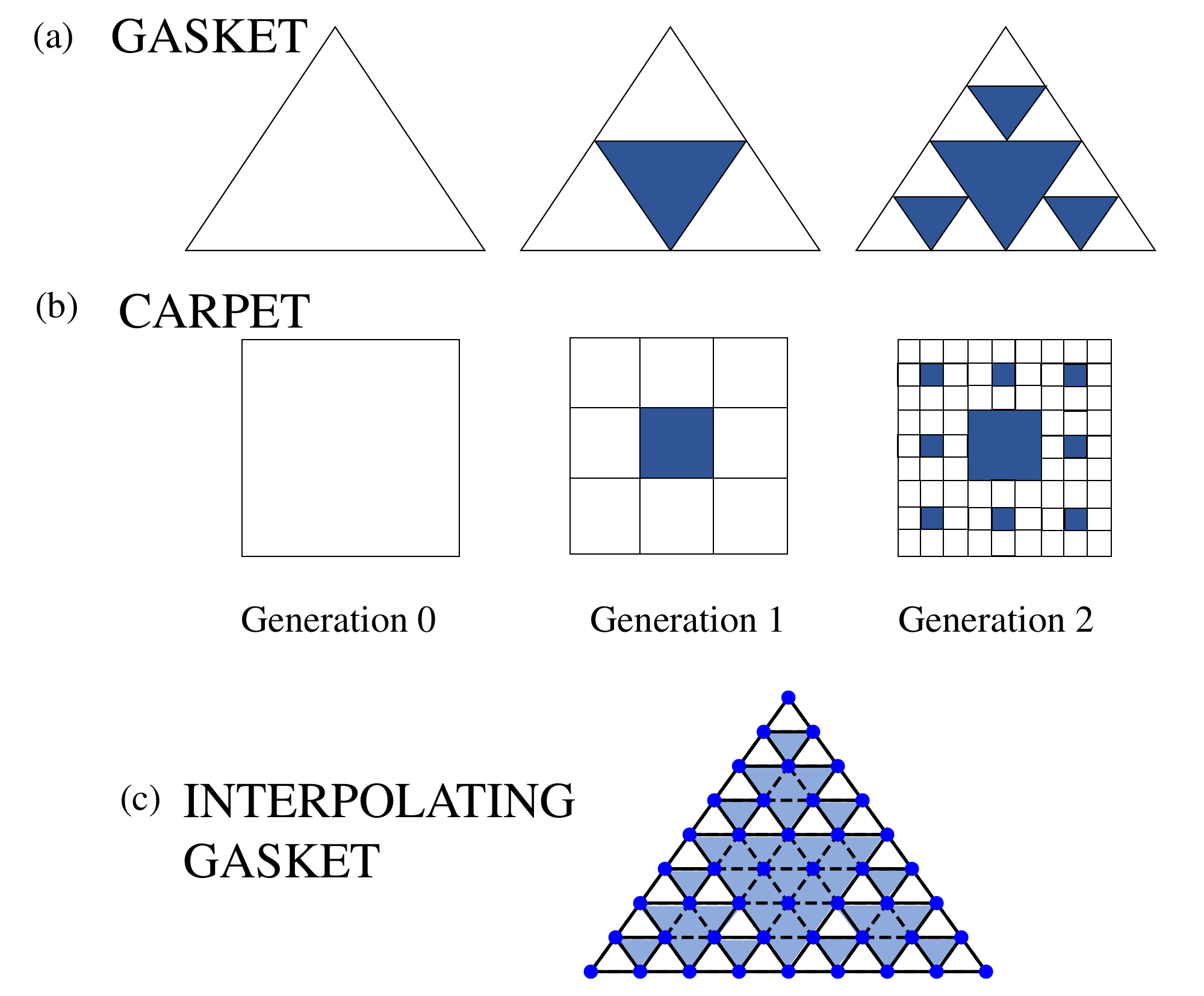}
    \caption{Construction scheme for (a) the Sierpiński gasket and (b) the Sierpiński carpet. In (c) we show a lattice, where the solid lines form a 3rd generation Sierpiński gasket, and the dashed lines interpolate between the fractal gasket and a regular triangular lattice.
    }
    \label{fig:fractals}
\end{figure}

Recent advances in the engineering of quantum systems have spurred quantum technology applications, including the vast field of quantum simulation. Different experimental platforms allow for the design and control of completely artificial quantum systems, with or without real-world counterpart. Recent examples for a simulation setup exploring the laws of quantum physics beyond standard geometries are quantum particles in fractal lattices, including electronic systems generated by molecular assembly \cite{Shang2015} or using scanning tunneling microscopy \cite{Kempkes_2019_b}, photonic systems of coupled optical fibers \cite{Xu_2021,Biesenthal_2022}, or cold atoms in optical tweezers \cite{Tian2023}. In general, fractal lattices are characterized by self-similar patterns repeated on different scales which give rise to a fractal Hausdorff dimension \cite{Mandelbrot_1967}. In the present article, we concentrate on Sierpiński fractals, specifically the Sierpiński gasket and the Sierpiński carpet. The self-similar construction scheme for these fractals is illustrated in Fig.~\ref{fig:fractals}(a,b). The fractal (Hausdorff) dimension of these structures is $d_f=\log(3)/\log(2) \approx 1.585$ for the gasket, and $d_f=\log(8)/\log(3)\approx 1.893$ for the carpet.

Exploring how the fractal geometry affects the dynamical behavior of quantum systems is an interesting research endeavor, and fascinating effects are found already in the single-particle domain: For instance, the combination of non-standard fractal geometry and topology has attracted significant interest \cite{Brzezinska2018,Pai2019,Iliasov2020,Fremling2020,Manna2022b}. The fate of topological edge states in fractal lattices, where a true bulk is absent, has now been studied experimentally using photonic waveguide arrays \cite{Biesenthal_2022}. 
Also in the absence of topological features, the transport in fractal lattices is a rich research subject. In general, transport behavior can be characterized through the mean square distance $\MSD(t)$ from the initial position, and in particular, through its scaling as a function of time:
\begin{align}
    \MSD(t) \sim t^{\alpha}.
\end{align} 
For $\alpha=1$, transport is diffusive, whereas for $\alpha=2$ transport is ballistic. Transport is sub-diffusive for $\alpha<1$, and hyper-ballistic for $\alpha>2$. The intermediate regime, $1<\alpha<2$, is called either super-diffusive or sub-ballistic.
Classical diffusion on fractals has been studied extensively since the 1980s \cite{Alexander_1982,Rammal_1983,Havlin_1987}, and sub-diffusive behavior with $\alpha=d_s/d_f$ has been established, where $d_s$ is the spectral dimension. In contrast to the fractal dimension $d_f$, the spectral dimension $d_s$ takes into account also the connectivity of the fractal lattice. It has a universal value, $d_s=4/3$, at percolation threshold according to the Alexander-Orbach conjecture \cite{Alexander_1982}. For Sierpiński fractals, the values $d_s = 2 \log(3)/\log(5) \approx 1.365$  and $d_s\approx 1.805$ have been obtained for gasket and carpet \cite{Darazs_2014}, respectively. 
Quantum-mechanical transport in the Sierpiński gasket has been contrasted to the classical random walk in Ref.~\onlinecite{Darazs_2014}. Studying the return probability of a quantum object evolving in the Sierpiński gasket, it has been shown that, instead of the classical decay $t^{-d_s/2}$, the quantum return probability in the Sierpiński gasket oscillates and remains above the classical value at all times. Notably, such a behavior is not apparent in the (finite-size) Sierpiński carpets, also studied in Ref.~\onlinecite{Darazs_2014}, hinting for different transport behavior of these two fractal structures. In Ref.~\onlinecite{vanVeen_2016}, quantum transport in Sierpiński carpets has been under scrutiny, also reporting clear differences between carpet and gasket. While in the Sierpiński gasket conductance is zero in extended energy regions, this is not the case in the Sierpiński carpet. As a possible geometric reason for this difference Ref.~\onlinecite{vanVeen_2016} mentions the infinite ramification number of the Sierpiński carpet \cite{Gefen_1984}, in contrast to a finite ramification number in the Sierpiński gasket. The ramification number counts the number of bonds that have to be cut in order to separate different iterations of the fractal.

The increased quantum return probability in the Sierpiński gasket can be seen as a dynamical consequence of the existence of localized eigenstates. Localized states in the Sierpiński gasket have first been found in Ref. \onlinecite{Domany_1983} using the Migdal-Kadanoff decimation technique. In fact, this early work had conjectured that \textit{all} quantum states in the Sierpiński gasket are exponentially localized, considering its spectral similarities to 1D quasi-crystals \cite{Aubry_1980,Kohmoto_1983}, and the fact that, like in disordered media or in quasi-crystals, the absence of Bloch's theorem can give rise to quantum interference effects which slow down the dynamics of a quantum object and possibly lead to Anderson localization \cite{Anderson_1958}. However, later work\cite{Wang_1995} has shown that the Sierpiński gasket exhibits a more complex behavior, as in addition to the localized states also an infinite number of extended states were found to live on the gasket. 
Recently, quantum transport in fractal geometries has been explored also experimentally in Ref.~\onlinecite{Xu_2021}, reporting super-diffusive quantum transport through Sierpiński gasket and carpet, with the scaling exponent $\alpha=d_f$ given by the fractal (Hausdorff) dimension of the lattice. 
Although these values are smaller than the ballistic diffusion exponent, $\alpha=2$, obtained for quantum diffusion on planar Bravais lattices \cite{Tang_2018,Razzoli_2020}, they would still constitute a significant quantum speed-up on fractals, in contrast to the increased return probability reported in Ref.~\cite{Darazs_2014} and the expected quantum localization effect. 

Given this controversial assessment on the transport behavior in Sierpiński fractals, the present manuscript revisits this scenario. For the gasket, we show that, in disagreement with the super-diffusive motion reported in Ref.~\cite{Xu_2021} and in line with the point of view of Anderson localization, the quantum particle behaves in a sub-diffusive way, with a quantum transport exponent $\alpha \approx 0.73$, smaller than the classical value $\alpha=d_s/d_f \approx 0.86$. 
On the other hand, for the carpet, our study confirms super-diffusive behavior with $\alpha\approx 1.8$. This surprising difference between the two structures can be understood from their different spectral properties, already noted in Ref.~\cite{Darazs_2014}. Quite generally it is known from random matrix theory that the spacing between adjacent energy levels provides deep insight into the dynamical behavior of a quantum system \cite{haake-book}. Specifically, ergodic systems exhibit level repulsion, and their level spacing distribution $p(s)$ has a power-law behavior $p(s)\sim s^\beta$ for $s\rightarrow 0$. This allows for classifying the system according to the exponent $\beta$. For the energy spectrum in fractal lattices, similarly to the case of quasi-crystals, level spacing analysis seems, on first sight, to be inappropriate, since the energy spectrum is characterized by huge degeneracies \cite{Pal2018}. However, in Refs.~\onlinecite{Geisel_1991,Sire_1993,Fleischmann_1995}, the concept of level spacing distribution has been adapted to the highly degenerate Cantor spectrum of quasi-periodic 1D models, and an inverse power law $p(s) \sim s^{-\beta}$ has been found. 
To test this behavior, the integrated level spacing distribution $p_{\rm int}(s) = \int_s^\infty ds' p(s')$ can be considered, by counting the number of gaps larger than $s$. In a finite system, this leads to a devil's staircase which, due to self-similarity of the  spectrum across various scales, can be smoothened to a power law, 
\begin{align}
   p_{\rm int}(s) \sim s^{1-\beta}.
\end{align}
 The exponent $\beta$ of the level spacing distribution and the exponent $\alpha$ of the mean square displacement can be related in the following way \cite{Geisel_1991,Fleischmann_1995}: By definition of the integrated level spacing distribution, the number of states which can be energetically resolved with an energy resolution $s$ (in units of the hopping energy $\hbar J$) is given by $p_{\rm int}(s) \sim s^{1-\beta}$. On the other hand, considering that the volume of a system scales with length $L$ (in units of the lattice constant $a$) as $L^{d_f}$, where $d_f$ is the Hausdorff dimension, we also have $L^{d_f} \sim p_{\rm int}(s)$. Hence, the smallest energy resolution $s$ is related to the length $L$ of the system as $L \sim  s^{(1-\beta)/d_f}$. At the same time, the relation $\MSD\sim t^\alpha$ connects a largest length scale $L$ to a largest time scale $t$ (in units $1/J$) via $L \sim t^{\alpha/2}$, or alternatively, to a smallest energy scale $s\sim t^{-1}$ via $L \sim s^{-\alpha/2}$. Combining these scaling relations leads to
\begin{align}
\alpha = \frac{2(\beta-1)}{d_f}.
\label{alphabeta}
\end{align}
As our numerical analysis of spectral behavior yields $\beta \approx 1.6$ on the gasket, Eq.~(\ref{alphabeta}) implies sub-diffusive behavior with $\alpha \approx 0.76$. On the other hand, on the Sierpiński carpet, we do not find an inverse power-law scaling of the level spacing distribution, which prevents us from applying  Eq.~(\ref{alphabeta}).

The stark contrast between sub-diffusive transport on the Sierpiński gasket and ballistic behavior on the regular lattice, together with the tunability of synthetic quantum lattices, opens an avenue to freely tune the transport behavior through all regimes by interpolating between the fractal lattice and the regular lattice, as illustrated in Fig.~\ref{fig:fractals}(c). In addition to this opportunity, we also discuss a possible application of the Sierpiński gasket as a quantum memory. Specifically, we demonstrate that the localized quantum dynamics on the Sierpiński gasket does not only significantly slow-down the spreading of a wave packet, but it also keeps memory of relatively fragile quantities like the phase of a quantum superposition. To this end, we compare the evolution of non-classical states, specifically symmetric and anti-symmetric arrangement of a de-localized object, and we find that the anti-symmetric superposition experiences slower initial spreading due to quantum interference. Strikingly, this leads to significantly different $\MSD(t)$ values even at long times, when in a regular lattice initial differences have been washed out. 

\begin{figure*}[tb]
    \centering
    \includegraphics[width=0.98\textwidth]{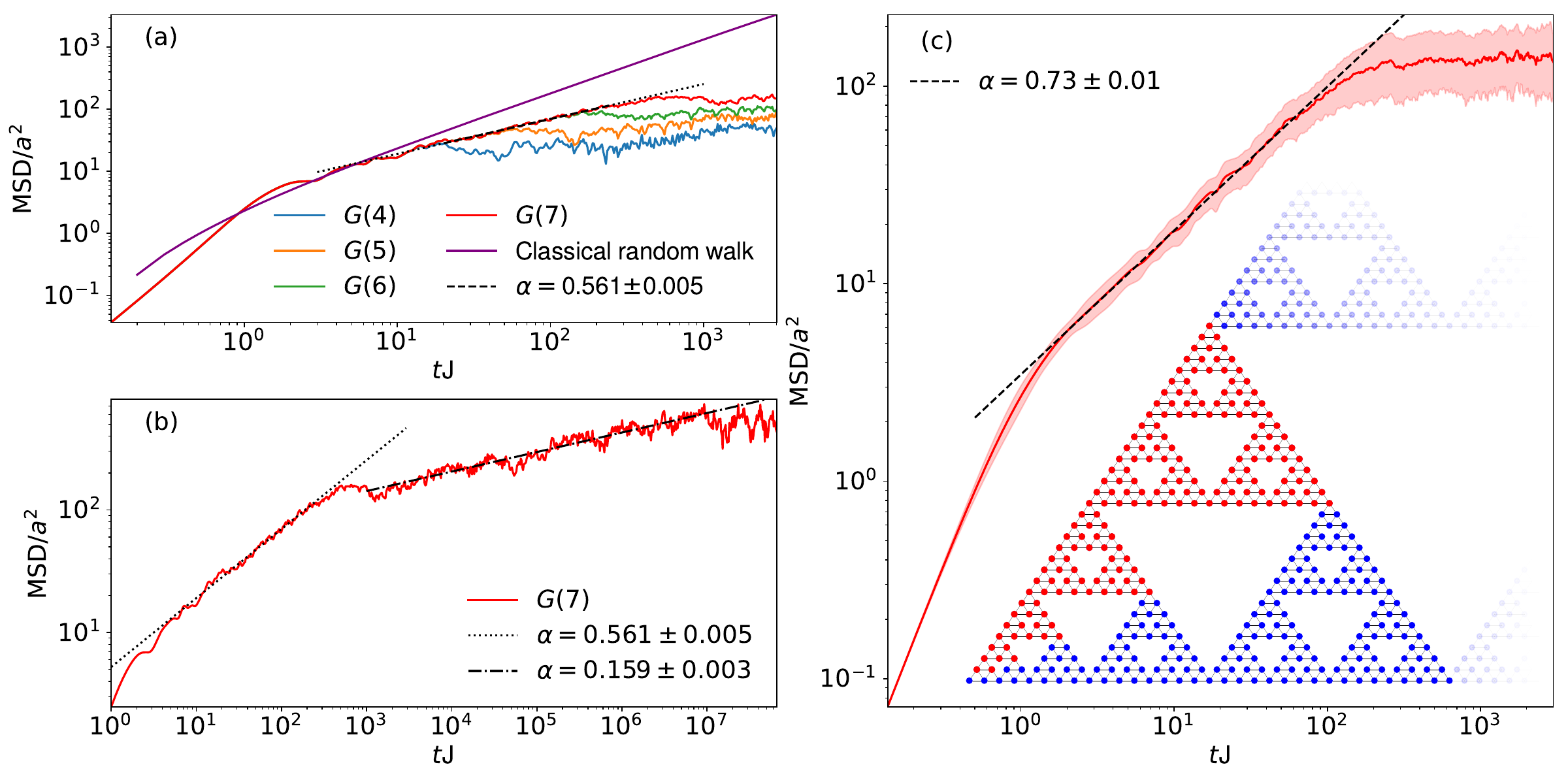}
    \caption{(a) Mean square distance $\MSD(t)$ of a particle starting from one corner of a Sierpiński gasket of generation $G(X)$ with $X=4,5,6,7$. This corresponds to $D=123,366,1095,3282$ sites, or triangles of length $L=16,32,64,128$. On time scales up to $t\sim1/J$, the particle behaves ballistically, $\MSD(t)\sim t^\alpha$ with $\alpha\approx 2.1$. On an intermediate time scale bound by the system size, the system evolves sub-diffusively, with an exponent $\alpha\approx0.56$. Beyond that regime, the $\MSD(t)$ almost flattens. The value reached at that time is still far away from the center of the triangle, $\MSD \ll L^2/3$. For comparison, we also plot the $\MSD$ of a classical continuous-time random walk on a G(7) gasket.
    (b) Same plot as in (a) for the 7th generation of the Sierpiński gasket, but on an extremely long time scale. We extract a long-time scaling $\alpha\approx 0.16$ from this figure.
    (c) For the 6th generation Sierpiński gasket, we analyze the behavior of $\MSD(t)$ averaged over a set of different initial positions, as indicated by the red colored sites, together with a standard deviation of the averaged $\MSD(t)$. In the intermediate temporal regime, we find $\alpha \approx 0.73$.
    }
    \label{fig:MSD_size}
\end{figure*}

\section{Results}

\subsection{Quantum transport on Sierpiński fractals}

We start by considering the mean square distance of a particle on the Sierpiński gasket which is initially prepared in one of the corners. For different generations of the fractal, the behavior is shown in Fig.~\ref{fig:MSD_size}(a). Each of these curves can be divided into three temporal regimes: 
\begin{itemize}
    \item Short times, $tJ \lesssim 1$: Ballistic regime. On short times, the system behaves ballistically, $\MSD(t)\sim t^\alpha$ with $\alpha\approx 2.1$. In this regime the system has yet no notion of the fractal geometry, and the behavior is the same as in a regular lattice. We note that the slightly hyper-ballistic value of $\alpha>2$ is a consequence of preparing the state near the boundary. For such initial conditions, also regular lattices exhibit the same increased value of $\alpha$.    
    \item Intermediate times, $ 1 \lesssim tJ \lesssim TJ$ with $T = (L/a)^{d_f}/(4J)$: Sub-diffusive regime. On intermediate-times, the system behaves sub-diffusively, with $\alpha \approx 0.56$. The extent of this regime is limited by the system size, determined by the fractal dimension $d_f$ and the side length $L$ of the triangle. We note again that also in this regime the exact value of $\alpha$ depends on the initial conditions, as we further discuss below. 
    \item Long times, $t \gtrsim T$: Quasi-localized regime. The evolution of the $\MSD(t)$ flattens further and becomes extremely slow. On long time scales, the behavior can be described on average with an exponent $\alpha \lesssim 0.15$, cf. Fig.~\ref{fig:MSD_size}(b).
\end{itemize}
It is important to note that at $t\approx T$, i.e. at the transition from the intermediate regime to the long-time regime, $\MSD(T)$ is still far below its thermalized value. In a thermalized system, the center of mass of the wave function would be at the center of the triangle, hence for a system initially prepared in one of the corners, the square distance between the corner and the center of the triangle defines the thermalized value, $\MSD_{\rm th} = L^2/3$. With the intermediate regime being too short to thermalize the system and with the subsequent evolution being extremely slow, it turns out that in the Sierpiński gasket the thermalized value will essentially never be reached. This can be seen from Fig.~\ref{fig:MSD_size}(b) which extends up to $tJ=10^6$, i.e. to times scales which are clearly beyond experimentally realistic values. Even on this time scale, the $\MSD$ remains below 1000 for a system with $L=128$, that is $\MSD_{\rm th} = 5461$. 
Of course, this example does not exclude the possibility of thermalization on even longer time scales which then become difficult to assess even in a numerical simulation due to the numerical precision. However, it is possible to argue rigorously that in the thermodynamic limit the system will not thermalize. Therefore, we note that $\MSD(T)$ scales sub-linearly with the system size, $\MSD(T)\sim T^\alpha \sim L^{0.56 d_f}$, in contrast to the quadratic size dependence of $\MSD_{\rm th}\sim L^2$. Hence, for larger systems the difference to a thermalized state gets more and more enhanced.

So far we have studied only the transport starting from a very special initial state where the particle is prepared in one corner of the triangle. However, in contrast to the case of an (infinite) Bravais lattice, the fractal lattice has non-equivalent lattice sites, and hence, the choice of initial state may affect the dynamical behavior. Indeed, when considering initial preparation on a variety of different sites, see Fig.~\ref{fig:MSD_size}(c), the exponent $\alpha$ in the intermediate regime tends to be larger for generic initial states as compared to an initial corner state. While the dynamics remains sub-diffusive for all initial states which we have considered, the $\MSD(t)$ averaged over different initial states, plotted in Fig.~\ref{fig:MSD_size}(c) together with its standard deviation, evolves with an exponent $\alpha \approx 0.73$.

The behavior on the Sierpiński gasket is in stark contrast to the quantum diffusion in a regular triangular structure. In that geometry, the ballistic initial behavior (with the possibility of $\alpha>2$ due to preparation near the boundary) is maintained up to a saturation time $T'\sim L^2$ at which the system enters a thermalized regime with $\MSD(t)$ oscillating around $\MSD_{\rm th}$. 

\begin{figure}
    \centering
    \includegraphics[width=0.49\textwidth]{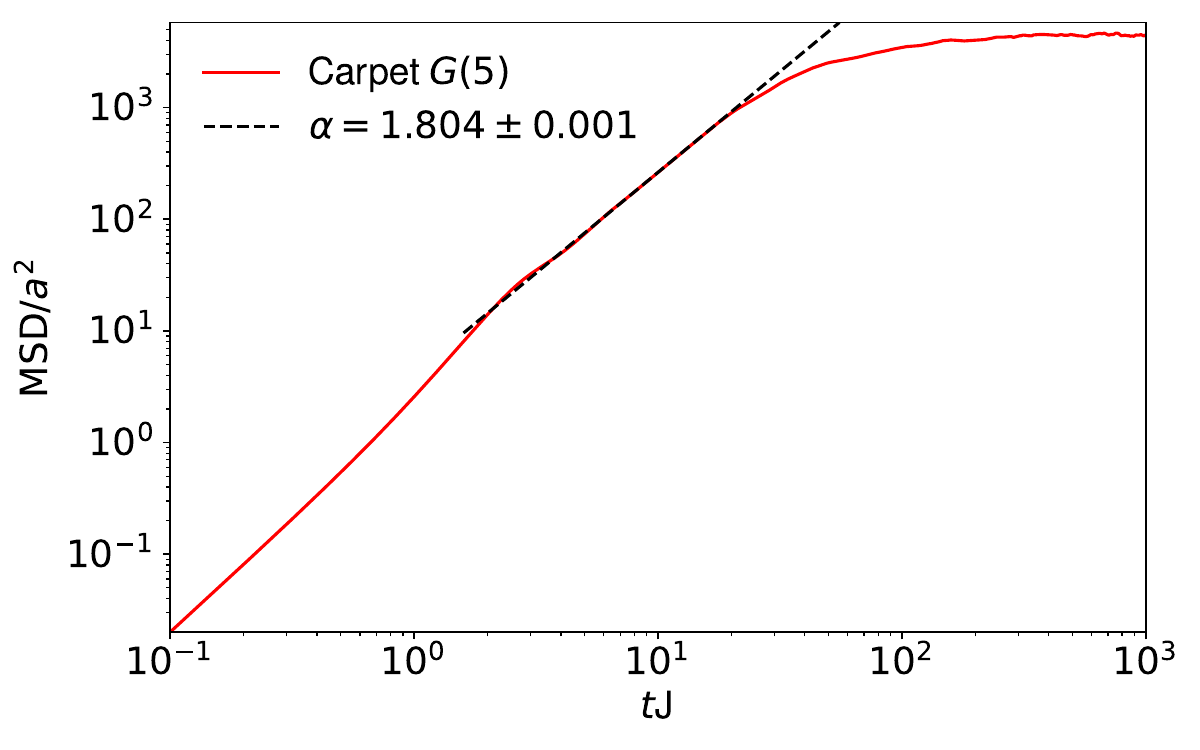}
    \caption{Average square distance as a function of time for a Sierpiński carpet lattice. The initial state is conformed by a particle on a corner of the lattice and evolves freely. There is a fit region with slope $\alpha\approx 1.80$.
    }
    \label{fig:sq_dist_carpet}
\end{figure}

The behavior on the Sierpiński gasket is also very different from the dynamics on the Sierpiński carpet. Instead of sub-diffusive transport, the carpet exhibits a highly super-diffusive, or sub-ballistic behavior, with $\alpha\approx 1.8$, see Fig.~\ref{fig:sq_dist_carpet}. This value is similar to the one previously obtained in Ref.~\cite{Xu_2021}, and it constitutes a significant quantum speed-up, compared to the classical value $\alpha=d_s/d_f\approx 0.95$. In this context, it should be noted that even for some non-Bravais periodic lattices sub-ballistic quantum transport has been found, e.g. with a value of $\alpha \approx 1.71$ for the honeycomb lattice \cite{Razzoli_2020}.

\subsection{Spectral properties on the Sierpiński gasket}

\begin{figure}
    \centering
   \includegraphics[width=0.48\textwidth]{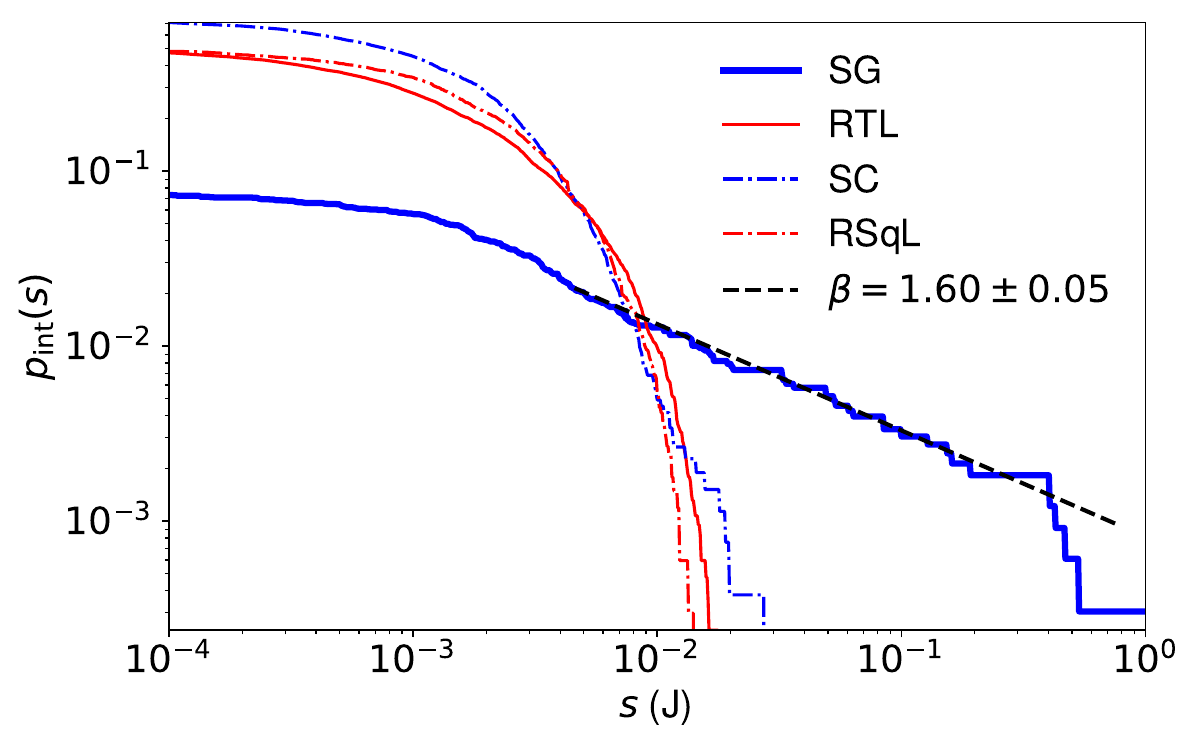}
    \caption{Integrated level spacing distribution $p_{\rm int}(s)$ for the energy spectra on different lattices: Solid lines are for triangular geometries, with the fractal  Sierpiński gasket of generation 7 (3282 sites) in the blue, and the corresponding regular triangular lattice (RTL) with 8385 sites in red. The dash-dotted lines correspond to square geometries, where the blue is the fractal Sierpiński carpet (SC) of generation 5 (5280 sites), and the red one to the regular square lattice (RSqL) with the same basis (6724 sites). Only on the gasket (SG), the integrated level spacing distribution exhibits an inverse power-law behavior $p_{\rm int}(s)\sim s^{1-\beta}$, which allows to determine the exponent $\beta$ via fitting, $\beta \approx 1.6$.
    }
    \label{fig:pint}
\end{figure}

We analyze the spectral properties of the system by plotting the integrated level spacing distribution, $p_{\rm int}(s)$, that is, the (normalized) number of energy gaps larger than $s$, see Fig.~\ref{fig:pint}. Within an extended region in energy, this staircase function is approximated by an inverse power law,  $p_{\rm int}(s)\sim s^{1-\beta}$,  as seen by using a double-logarithmic axis scale. Numerically, we obtain $\beta\approx 1.6$. The proximity of $\beta$ to the Hausdorff $d_f$ seems suggestive that both quantities might be identical, but we are lacking any \textit{a priori} argument for such a relation. However, according to the argumentation outlined in Sec.~\ref{sec:intro}, the fractal dimension relates $\beta$ to $\alpha$ through Eq.~(\ref{alphabeta}). Considering a finite precision $\beta=1.6 \pm 0.05$, we expect $\alpha=0.76\pm 0.06$, in accordance with the $\alpha$ obtained before by averaging over different initial states.

Apart from that quantitative agreement between the exponent characterizing level spacing, $\beta$, and the exponent characterizing transport, $\alpha$, the integrated level spacing distribution also demonstrates an interesting qualitative difference between the Sierpiński gasket and other structures. Neither a regular lattice with triangular or square geometry, nor the Sierpiński carpet exhibits an extended spectral regime which can be approximated by an inverse power-law, see Fig.~\ref{fig:pint}. As we have argued above, both the Sierpiński carpet and regular lattices (square of triangular) exhibit much faster transport behavior than the Sierpiński gasket, within or close to the ballistic regime.

\subsection{Transport on interpolating lattices}

\begin{figure*}
    \centering
    \includegraphics[width=0.96\textwidth]{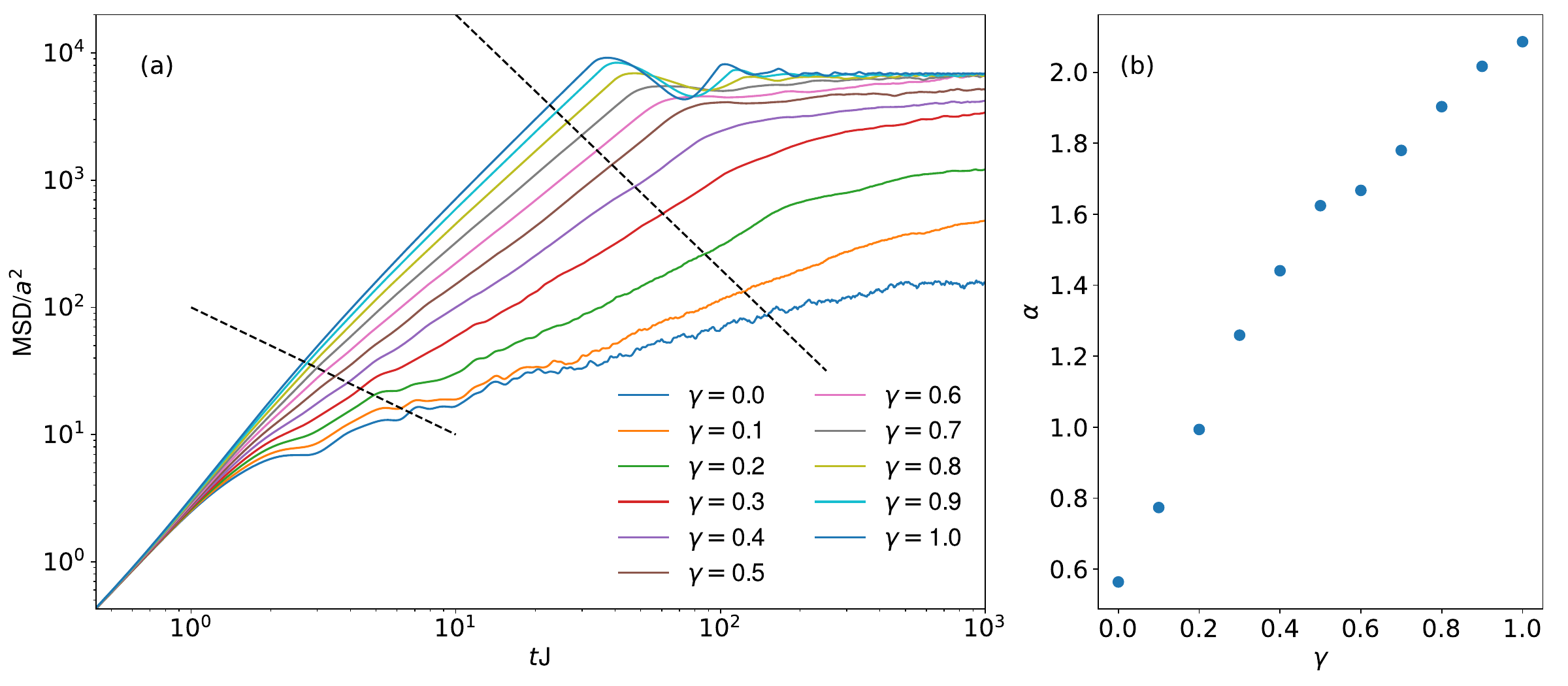}
    \caption{(a) Mean square distance $\MSD(t)$ of a particle in an interpolating lattice, cf. Fig.~\ref{fig:fractals}(c), characterized by the ratio $\gamma\equiv J'/J$ between hopping parameters $J'$ exclusive to the regular lattice, and $J$ in both regular and fractal lattice. We initialize the evolution in one corner of an interpolating gasket of generation 7.  The slowest behavior is obtained in a fully fractal geometry ($\gamma=0$), whereas the fastest behavior corresponds to regular triangular lattice ($\gamma=1$). (b) For the different values of $\gamma$, we extract the exponent $\alpha$ of the mean square distance (from fits to the curves in (a) in the intermediate regime marked by the dashed lines). The result is plotted as a function of $\gamma$.
    }
    \label{fig:inter}
\end{figure*}

The very different transport behavior of Sierpiński gasket and regular triangular lattice open up a route to tailor-made transport behavior by interpolating between these two cases, as sketched in Fig.~\ref{fig:fractals}(c). The interpolating lattice contains all sites of the regular lattice, but for the bonds at those sites which are exclusive to the regular lattice a different hopping amplitude $J'$ is chosen (as compared to the hopping amplitude $J$ on the fractal).  In Fig.~\ref{fig:inter}(a), the $\MSD(t)$ is plotted for various interpolating choices $\gamma\equiv J'/J$.  Strikingly, also in the intermediate case (i.e. $0<J'<J$), the transport behavior can be separated into three temporal regimes. Our main interest is the exponent $\alpha$ for the intermediate temporal regime, in between the dashed lines of Fig.~\ref{fig:inter}(a). We plot this value $\alpha$ in Fig.~\ref{fig:inter}(b), showing that the transport properties can continuously be tuned from the sub-diffusive regime in the fractal lattice $\gamma \lesssim 0.3$), through a super-diffusive regime ($0.3 \lesssim \gamma \lesssim 0.8$), into a ballistic regime in the (almost) regular lattice ($\gamma \gtrsim 0.8$).

\section{Discussion}
\begin{figure*}
    \centering
    \includegraphics[width=0.96\textwidth]{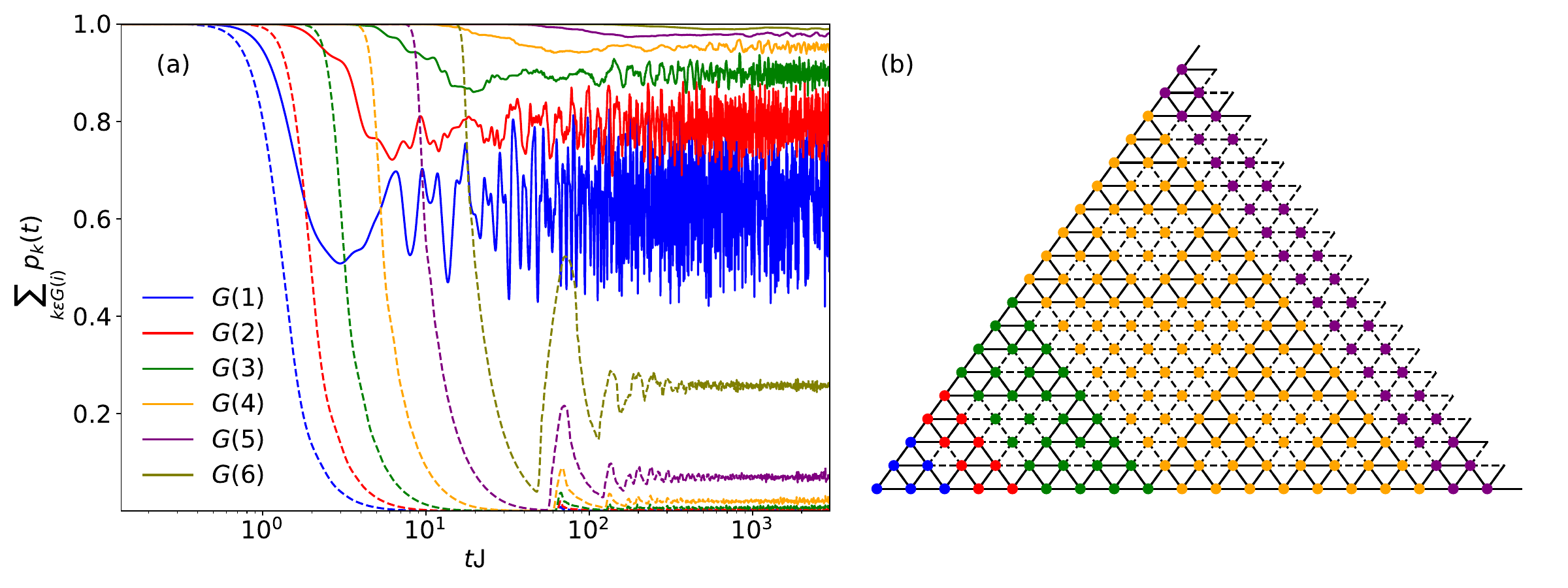}
    \caption{After preparing the system initially in the lower left corner of a G(7) Sierpiński gasket (or the corresponding regular triangle, we plot the weight of the wave function within the lower-leftmost G($i$) structure (as indicated in the right panel). In the regular lattice the weight decays quickly and thermalizes at the thin dashed lines, corresponding to the ratio of site numbers $N_s[G(i)]/N_s[(G7)]$. In contrast, the weight in the fractal, will always keep some memory of the initial state. 
    }
    \label{fig:memory}
\end{figure*}

Our results have established that the dynamics on the Sierpiński gasket is coined by the localized eigenstates and an inverse-power-law level spacing distribution, in stark contrast to the case of regular lattices or Sierpiński carpet. We now discuss how this localized nature of the gasket might be used as a quantum memory. 
Clearly, the slow growth of the $\MSD(t)$ and the demonstrated inability of reaching a thermalized value keep memory of the classical information about the initial position of the particle. 
This is also illustrated in Fig.~\ref{fig:memory}, showing that after initial preparation in the corner of a G(7) gasket, the weight of the time-evolved wave function will remain concentrated in the surrounding G(1) gasket (indicated in blue). Considering the surrounding G(6) structure, i.e. roughly 1/3 of the total lattice, this will keep more than 95 \% of the weight for all times. In contrast, for the case of a regular lattice we see a rapid drop to the thermalized value 1/3.

\begin{figure}
    \centering
    \includegraphics[width=0.48\textwidth]{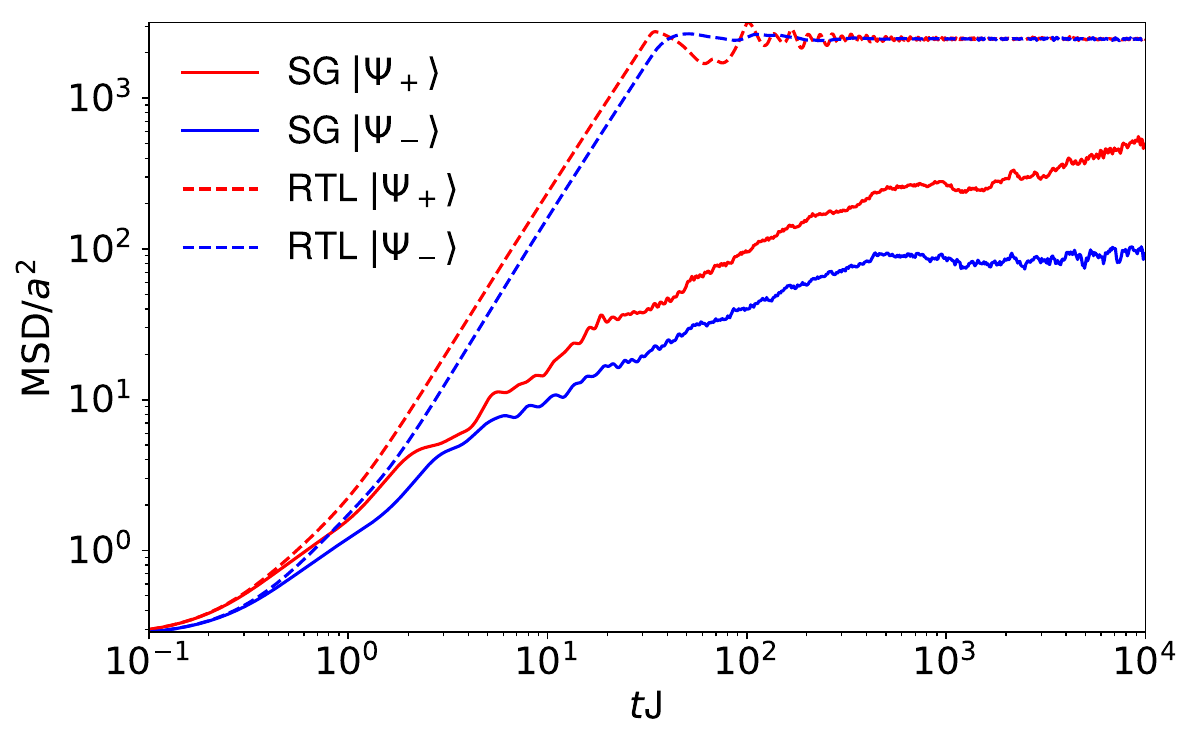}
    \caption{Mean square distance $\MSD(t)$ of a particle with a non-classical initial state. The particle is prepared in a superposition of being in the corner and one of its first neighbors. The solid red line corresponds to a particle in a G(7) Sierpiński gasket with a symmetric initial configuration. The solid blue line corresponds to the anti-symmetric initial condition for the fractal geometry. The dashed lines correspond to a standard triangular lattice geometry with the same basis as the fractal considered. }
    \label{fig:memory2}
\end{figure}

Importantly, the Sierpiński gasket is also able to memorize quantum properties of the initial state. To this end, we consider the initial quantum superposition
\begin{align}
|\Psi_{\pm} \rangle = \frac{1}{2} \left( |A\rangle \pm |B\rangle \right),
\end{align}
with $+$ denoting the symmetric, and $-$ the anti-symmetric superposition. The states $|A\rangle $ and $|B\rangle $ denote two different initial positions, where for concreteness we choose  $|A\rangle $ to be a corner state and $|B\rangle $ its neighbor. Defining $\MSD(t)$ with respect to their center-of-mass, we find that the anti-symmetric state $|\Psi_{-} \rangle$ evolves slower as compared to the symmetric state $|\Psi_{+} \rangle$, see Fig.~\ref{fig:memory2}. We attribute this difference to destructive interference effects during the simultaneous tunneling from $A$ and $B$ to their common neighbor. Such a confinement effect stemming from the phase of the wave function is also found initially on a regular lattice. However, on longer time scales only the fractal lattice keeps memory of the initial phase difference in form of a significantly different $\MSD(t)$. In the regular lattice, as can also be seen from Fig.~\ref{fig:memory2}, both initial states evolve to the same $\MSD_{\rm th}$, and there is no obvious indicator of the initial phase difference.

In future work, it will be interesting to explore this effect beyond single-particle physics. For example, one could consider two or more entangled particle evolving quantum-dynamically but under the influence of a certain measurement rate. We expect that the measurement-induced entanglement transition \cite{Skinner2019} will depend on the geometry, and the Sierpiński gasket will maintain the entanglement at higher measurement rates as compared to regular lattices or the Sierpiński carpet. In the many-body regime, we expect to find a glass and/or many-body localized phase in the Sierpiński gasket, whereas such a phase is not expected on the carpet. In view of the computational complexity of quantum many-body physics and open quantum systems, we expect that quantum simulations with interacting particles on fractal lattices will be particularly useful and provide important new insights into exotic quantum phenomena. This may include electronic and atomic fractal systems, cf. Refs.~\cite{Shang2015,Kempkes_2019_b,Tian2023}, or by adding optical non-linearities to the photonic simulations. 
So far, theoretical attempts to study quantum many-body phases in fractal lattices include studies of quantum phase transitions and quantum criticality in interacting spin models \cite{Yi2015,Xu2017,Krcmar2018}, the study of interacting topological systems, in particular with respect to the fate of anyons \cite{Manna2020,Manna2022,Li2022}, or the very recent mean-field study of the Bose-Hubbard model on the Sierpiński gasket \cite{Koch2024}.

\section{Methods}
\subsection{Quantum transport}
We study tight-binding systems described by a Hamiltonian of the form
\begin{align}
H/\hbar = -\sum_{i,j} J_{i,j} a_i^\dagger a_j + \sum_i \epsilon_i n_i.
\end{align}
Our focus is on fractal lattices, in particular the Sierpiński gasket and Sierpiński carpet, where sites $i$ are the vertices of the structure. The construction scheme for these fractals is illustrated in Fig.~\ref{fig:fractals}(a,b).
The tunneling amplitude $J_{i,j}=J \delta_{\langle i,j\rangle}$ is non-zero between nearest-neighbors. We will also study the case of an interpolating lattice, as shown in Fig.~\ref{fig:fractals}(c), where we have nearest-neighbor hopping on a regular lattice, but with two types of couplings, $J$ belonging to the Sierpiński fractal, and $J'$ for the others. The on-site frequencies $\epsilon_i$ are, where not otherwise defined, homogeneous, $\epsilon_i=\epsilon$. With this choice, the diagonal term of the Hamiltonian is proportional to the identity matrix and only contributes an irrelevant overall phase factor. Hence, we choose $\epsilon=0$. By numerical diagonalization of $(H/\hbar)$, we find the eigenvectors $| \alpha \rangle$ and eigenvalues $\omega_\alpha$ of the tight-binding model on finite lattices, which then allows us to evolve an arbitrary initial state $|\Psi(0)\rangle$ to time $t$,
\begin{align}
|\Psi(t)\rangle = \sum_\alpha  \langle \alpha | \Psi(0) \rangle e^{-i \omega_\alpha t} |\alpha\rangle.
\end{align}
We are then interested in different observables which are best defined in a local basis $|i \rangle = a_i^\dagger |{\rm vac}\rangle$, that is, a basis of states where the particle exclusively occupies one site $i$. Specifically, 
the probability to be at a given site $i$
at time $t$ reads
$p_i(t) = |\langle i | \Psi(t) \rangle|^2$. If the particle has initially been prepared at a site $i$, i.e. $|\langle i |\Psi(0)\rangle|=1$, the quantity $p_i(t)$ equals the return probability of the quantum walk. Another interesting quantity is the mean square distance $\MSD(t)$. Let again be $|\langle i |\Psi(0)\rangle|=1$, and let ${\bf r}_j$ denote the Euclidean coordinates at any site $j$.  The mean square distance is then defined as
\begin{align}
\MSD(t) = \sum_j |{\bf r}_j-{\bf r}_i|^2 |\langle j| \Psi(t) \rangle |^2.
\label{eq:MSD}
\end{align}

\subsection{Continuous-time classical random walk}
With a proper choice of the on-site potentials $\epsilon_i$, the Hamiltonian $H$ also defines 
an analog classical evolution, cf. Ref.~\onlinecite{Farhi_1998}. In the classical random walk, the probability of moving from site $i$ to site $j$ during a small time interval $\tau$ is given by $-\tau \langle j |H| i \rangle = \tau J$, for connected sites $i$ and $j$. If $i$ is connected to $N_i$ different sites, the total probability of a move is $\tau J N_i$. The probability of remaining on the site shall be given by $1-\tau \langle i | H  | i \rangle = 1- \tau \epsilon_i$. To keep the probability normalized, we must have $\epsilon_i = N_i J$. On a Sierpiński gasket, $\epsilon_i= 4J$ for all sites, except for the three corner sites, where we have $\epsilon_i=2J$.
The definition of probabilities after an infinitesimal time step $\tau$ defines the probabilities for all times through a Schr\"odinger-like equation $\frac{\rm d}{{\rm d}t} p_{ji}(t) = - \sum_k \langle j | H |k\rangle p_{ki}(t).$ Under the boundary condition $p_{ji}(0)=\delta_{ji}$, with $i$ denoting the site of initial preparation, the differential equation is solved by $p_{ji}(t) =\langle j | e^{-Ht} | i \rangle$. From this, we define the classical return probability $p_{ii}(t)$, or the mean square distance of the classical diffusive process by replacing $|\langle j| \Psi(t) \rangle |^2$ in Eq.~(\ref{eq:MSD}) by $p_{ji}(t)$.

\acknowledgments
T.G. acknowledges funding by Gipuzkoa Provincial Council (QUAN-000021-01), by the Department of Education of the Basque Government through the IKUR strategy and through the project PIBA\_2023\_1\_0021 (TENINT), by the Agencia Estatal de Investigación (AEI) through Proyectos de Generación de Conocimiento PID2022-142308NA-I00 (EXQUSMI), by the BBVA Foundation (Beca Leonardo a Investigadores en Física 2023).
The BBVA Foundation is not responsible for the opinions, comments and contents included in the project and/or the results derived therefrom, which are the total and absolute responsibility of the authors.
B.J-D and A.R-F acknowledge funding from Grant No.~PID2020-114626GB-I00 by MCIN/AEI/10.13039/5011 00011033 and 
"Unit of Excellence Mar\'ia de Maeztu 2020-2023” 
award to the Institute of Cosmos Sciences, Grant CEX2019-000918-M funded by MCIN/AEI/10.13039/501100011033. 
We acknowledge financial support from the Generalitat de Catalunya (Grant 2021SGR01095). 
A.R.-F. acknowledges funding from MIU through Grant No. FPU20/06174.
U.B. acknowledges support from: ERC AdG NOQIA; MCIN/AEI (PGC2018-0910.13039/501100011033,
CEX2019-000910-S/10.13039/501100011033, Plan National FIDEUA PID2019-106901GB-I00, Plan National
STAMEENA PID2022-139099NB-I00 project funded
by MCIN/AEI/10.13039/501100011033 and by the
“European Union NextGenerationEU/PRTR” (PRTRC17.I1), FPI); QUANTERA MAQS PCI2019-111828-
2); QUANTERA DYNAMITE PCI2022-132919 (QuantERA II Programme co-funded by European Union’s
Horizon 2020 program under Grant Agreement No
101017733), Ministry of Economic Affairs and Digital Transformation of the Spanish Government through
the QUANTUM ENIA project call – Quantum Spain
project, and by the European Union through the Recovery, Transformation, and Resilience Plan – NextGenerationEU within the framework of the Digital Spain 2026
Agenda; Fundaci´o Cellex; Fundaci´o Mir-Puig; Generalitat de Catalunya (European Social Fund FEDER and
CERCA program, AGAUR Grant No. 2021 SGR 01452,
QuantumCAT U16-011424, co-funded by ERDF Operational Program of Catalonia 2014-2020); Barcelona
Supercomputing Center MareNostrum (FI-2023-1-0013);
EU Quantum Flagship (PASQuanS2.1, 101113690);
EU Horizon 2020 FET-OPEN OPTOlogic (Grant No
899794); EU Horizon Europe Program (Grant Agreement
101080086 — NeQST), ICFO Internal “QuantumGaudi”
project; European Union’s Horizon 2020 program under
the Marie Sklodowska-Curie grant agreement No 847648;
“La Caixa” Junior Leaders fellowships, La Caixa” Foundation (ID 100010434): CF/BQ/PR23/11980043. Views
and opinions expressed are, however, those of the author(s) only and do not necessarily reflect those of the European Union, European Commission, European Climate, Infrastructure and Environment Executive Agency
(CINEA), or any other granting authority. Neither the
European Union nor any granting authority can be held
responsible for them. 
U.B. is also grateful for the financial support of the IBM Quantum Researcher Program.

\section*{Contributions}
T.G. conceived and supervised the project. A.R.F., P.P., and T.G. developed the codes and performed numerical simulations. A.R.F., B.J.-D., T.G., and U.B.  analyzed and interpreted the data. A.R.F. and T.G. wrote the manuscript with the feedback from all co-authors.

\bibliography{bib}
\end{document}